\documentclass[structabstract,printer]{aa} % for a referee version
\usepackage{graphicx}
\usepackage{natbib}
\usepackage{amsmath}
\usepackage{txfonts}
\usepackage{color}
\usepackage{enumerate}

\newcommand{\be}{\begin{eqnarray}}
\newcommand{\ee}{\end{eqnarray}}

\newcommand {\nbodypp}{\textsc{\mbox{nbody6\raise.4ex\hbox{\tiny++}}}}
\newcommand {\Msun} {\mbox{M$_{\odot}$}}

\begin{document}

\title{Observational constraints on star cluster formation theory -\\ I. The mass-radius relation}

\authorrunning{Pfalzner et al.}
\titlerunning{The mass-radius relation}

\author{S. Pfalzner\inst{1}, H. Kirk\inst{2}, A. Sills\inst{3}, J. S. Urquhart\inst{1,4}, J. Kauffmann\inst{1}, M. A. Kuhn\inst{5,6}, A. Bhandare\inst{1}, K. M. Menten\inst{1}}
\institute{
\inst{1}Max-Planck-Institut f\"ur Radioastronomie, Auf dem H\"ugel 69, 53121 Bonn, Germany\\
\inst{2}National Research Council Canada, Herzberg Astronomy and Astrophysics, Victoria, BC, V9E 2E7, Canada\\
\inst{3}Department of Physics and Astronomy, McMaster University, 1280 Main St. W., Hamilton, ON L8S 4M1, Canada\\
\inst{4}Centre for Astrophysics and Planetary Science, University of Kent, Canterbury, CT2 7NH\\
\inst{5}Instituto de Fisica y Astronom\'ia, Universidad de Valpara\'iso, Gran Breta\~na 1111, Playa Ancha, Valpara\'iso, Chile\\
\inst{6}Millennium Institute of Astrophysics, Santiago, Chile\\
\email{spfalzner@mpifr.de}}
\date{ }

\abstract
{Stars form predominantly in groups usually denoted as clusters or associations. The observed stellar groups display a broad spectrum of masses, sizes and other properties, so it is often assumed that there is no underlying structure in this diversity.}
{Here we show that the assumption of an unstructured multitude of cluster or association types might be misleading. Current data compilations of clusters in the solar neighbourhood show correlations between cluster mass, size, age, maximum stellar mass etc. In this first paper  we take a closer look at the correlation of cluster mass and radius.}
{We use literature data to explore relations in cluster and molecular core properties in the solar neighbourhood.   }
{We show that for embedded clusters in the solar neighbourhood there exists a clear correlation between cluster mass and half-mass radius of the form $M_c = C R_c^\gamma$ with $\gamma$= 1.7 $\pm$0.2. This correlation holds for infra red $K$ band data as well as X-ray sources and for clusters containing a hundred stars up to those consisting of a few tens of thousands of stars. The correlation is difficult to verify for clusters containing $<$30 stars due to low-number statistics.
Dense clumps of gas are the progenitors of the embedded clusters. We find a similar slope for the mass--size relation of dense, massive clumps as for the embedded star clusters. This might point at a direct translation from gas to stellar mass: however,  it is difficult to relate size measurements for clusters (stars) to those for gas profiles. Taking into account multiple paths for clump mass into cluster mass, we obtain an average star-formation efficiency of 18\%$^{+9.3}_{-5.7}$ for the embedded clusters in the solar neighbourhood.}
{The derived mass--radius relation gives constraints for the theory of clustered star formation. Analytical models and simulations of clustered star formation have to reproduce this relation in order to be realistic.}

\maketitle
\section{Introduction}
An important unanswered question in modern astrophysics concerns the mechanism(s) by which star clusters form. For more than ten years it has been known that most stars form in such clusters (Lada \& Lada 2003, Porras et al. 2003) which vary widely in their properties, like the numbers of stars they contain or their physical sizes. For example, one finds sparsely populated clusters, e.g. in Taurus (Kirk \& Myers (2011), as well as very dense systems containing tens of thousands of stars like Arches (for example, Stolte et al. 2015). Given this multitude of clusters, the question is whether any underlying relations provide the recipe for clustered star formation. 

In recent years, many new clusters have been discovered and more detailed information has been collected for those already known. At the same time, simulations of clustered star formation are incrementally advancing through inclusion of more physical processes like magnetic fields and feedback (e.g., Tilley \& Pudritz 2004, Bate 2009, Offner et al. 2009, Federarth et al. 2010, Dale et al. 2012). Here, the main problem is that these ab initio calculations are subjected to computational constraints which limit the investigations to at most a few hundred stars in the system. Various analytic models have also been used to describe the formation of massive clusters (for example, Elmegreen 2008, Dib et al. 2013, Parmentier \& Pfalzner 2013).

\begin{figure*}[t]
\includegraphics[width=1.1\textwidth]{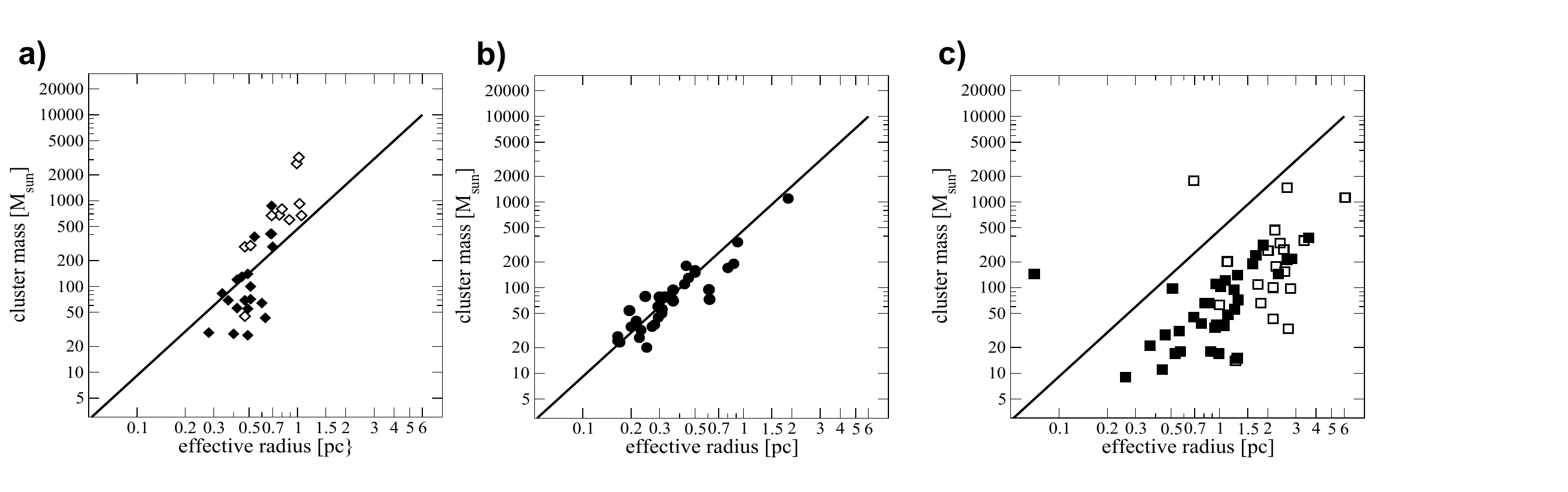}
\caption{Cluster mass as a function of effective cluster radius for the data from a) Carpenter 2000 (diamonds), b) Lada \& Lada (2003) (circles) and c) Kumar et al. (2006) (squares). The line shows the fit by Pfalzner (2011) for the
Lada\& Lada (2003) data. Clusters for which holds \mbox{$\bar{m}$=0.4 - 0.6\,\Msun } $M_C/N$ are indicated by full sybols, whereas open symbols indicate those that do not fulfil this criterion. }
\label{fig:cluster_mass_radius}
\end{figure*}

The comparison between simulation/theory and observations is often driven by visual similarities. So has the observational finding of extended filamentary structures (for a review see Andre et al. 2013) seeded a renewed interest in
the effect of sub-clustering and mass segregation, which has led to an extensive debate on their relative importance in different theoretical and numerical models. Many of these issues concern the initial condition for star cluster formation, like whether clusters preferentially form from a spherical or elongated gas clump (Myers 2011) and whether the final clusters forms as a single entity
(Banerjee \& Kroupa 2014, 2015) or as a merger product from a sub-clustered structure (McMillan 2007, Allison et al. 2009, Maschberger \& Clarke 2011, Moeckel \& Bonnell 2009, Girichidis et al. 2012). All these questions are very important for a final understanding of star formation.  However, over the last decade observations have revealed a number of correlations between global cluster properties, which have received less attention from theory. These concern relations that describe either the temporal development of clusters
or statistics of the diversity of fully formed clusters. What is missing is a thorough quantitative comparison between the numerical and analytical models with these recent observational results.

The observed diversity in clusters is often implicitly interpreted as an absence of  correlations between cluster properties. In other words, it is assumed that - within a very broad spectrum - any type of cluster can form. Collecting data from different types of observations, we will show that, in reality, the parameter space is much more restricted. Star-cluster formation occurs with well defined property correlations. Thus, cluster formation theory must not only get the initial conditions right, but must also fulfil all of these property constraints simultaneously.

 As a first step towards this goal, we provide here and in an accompanying paper a list of constraints on star-formation theory that can be provided by observations. We will discuss these
constraints one-by-one and point out the different interpretation of these observational facts.
In an accompanying paper, we describe the constraints given by  observations of the local star formation efficiency  $\epsilon_{SFE}$, the mass of the most massive star $m_{max}$, and the cluster expansion history. 

In this first paper, we put special emphasis on the relation between cluster mass $M_c$ and size $R_c$, often expressed in terms of the cluster half-mass radius $r_{hm}$, in the embedded phase, as many new data on this relation emerged recently.  We will compare different observational results and try to understand the connection to the star-formation process. This will be done for different star-forming regions, but with an emphasis on the solar neighbourhood, where particularly detailed observations are available. We explicitly exclude  the Galactic Centre region and the high-stellar-density areas in the spiral arms from this study, because in such extreme environments, clustered star formation might proceed in a different way (Pfalzner 2009, Longmore et al. 2013, Pfalzner \& Kaczmarek 2013, Kruijssen et al. 2014). An indication that this might be the case is the presence of massive compact clusters like Arches and Quintuplet in the Galactic Centre, which do not have counterparts in the solar neighbourhood. Equivalently dense systems  are found in the spiral arms (NGC 3603, Westerlund 1 and 2, etc.).

In Section\,2, we take a closer look at the $M_c-R_c$ relation in $K$ band and X-ray data for clusters that typically contain one to several hundred stars. We extend the investiagtion to high- and low-mass clusters and discuss the influence of the definitions of masses and radii for clusters. In Section\,3, we investigate  the precursors of clusters - the massive gas clumps in molecular clouds. This is followed by a comparison of the different cluster formation models and simulations of observational constraints in Section\,4. In Section\,5, the possible origin of the mass-radius relation is discussed and the differences in the Galactic Centre environment pointed out.   

\section{The $M_c/R_c$ relation}

\subsection{$K$-band observations} 

Here we want to investigate whether a relation between the mass and size of young clusters ($<$ 4\,Myr) exists. At such young ages most clusters are still embedded in their natal gas, apart from the most massive clusters where gas expulsion happens more rapidly (1- 2\,Myr), presumably due to their high numbers of massive stars. 

The first compilations of young-cluster properties, important in this context, were those of 
Carpenter (2000) and Lada \& Lada (2003). Carpenter used the 2MASS Second Incremental Release Point Source Catalog to identify clusters by investigating the spatial distribution in $K_s$ band density maps of young stars in the Perseus, Orion A, Orion B, and Mon R2 molecular clouds. He subtracted a semi-empirical model of the field star contamination from the observed star counts and used stellar surface density maps to identify compact clusters and stellar populations more uniformly distributed over the molecular cloud. His sensitivity calculations suggest that the number of stars in the distributed population may be underestimated by a factor of 2 or more. 

Lada \& Lada (2003) concentrated on embedded clusters within $\sim$2\,kpc of the Sun because such a sample is the most statistically complete with detailed observational data being available.
They consider clusters to be groups of stars that contain at least 35 members and have observed stellar-mass volume density exceeding $\rho_s >$ 1.0\,\Msun\ pc$^{-3}$. They searched publications between 1988 and 2003, and, as a result, the methods employed to determine the cluster properties are not strictly uniform.
 
Figure\,1 shows observed cluster masses and cluster radii for the data compilations of a) Carpenter et al (2000), and b) Lada \& Lada (2003). Combining the two data sets, Adams et al. (2006)
found that the number of stars $N$ and the cluster radius $R_c$ are roughly correlated as
\be R_c (N) = C_R \sqrt{(N/300)}\ee
where $C_R  \approx$ 1-2\,pc. If we assume a mean stellar mass \mbox{$\bar{m}$=0.5 - 0.6\,\Msun } and use $M_c = N \bar{m}$, Eq.\,1 corresponds to
\be M_c = C_M R_c^2\ee
where $C_M$= 75 - 180. In the following we look at cluster data that have been published since 2003 to test whether they obey similar relations of the more general form
\be M_c = C_M R_c^\gamma.\ee
Pfalzner (2011) fitted the mass and radius data from Lada \& Lada (2003) only (see Fig.\,1b),  and obtained
\be M_c = 359 R_c^{1.71 \pm 0.07}.\ee
This line is shown in all three plots of Fig.\,1 to provide a comparison. 

The reason for the greater power-law index in Eq.\,2 in comparison to Eq.\,4 is that Pfalzner did not include the data by Carpenter et al. (2000), which have a  steeper slope (see Fig.1a) of $\gamma \approx$ 3.2). However, there is a problem with the data compilation of Carpenter et al. (2000).  
If the initial mass function (IMF) were ideally sampled over a stellar mass range \mbox{(0.08--160\,\Msun),} one would expect a relation between the cluster mass and the average stellar mass $\bar{m}$ of the form $M_c = N \bar{m}$, where $\bar{m}$ is typically 0.5-0.6\,\Msun . However, probably due to observational limits, some clusters in the Carpenter sample deviate significantly from   $\bar{m}$=0.5 - 0.6\,\Msun, and contain significantly more low- or high-mass stars. These clusters are indicated as open symbols in Fig.\,1a. Obviously the high-mass clusters in this sample are dominated by clusters for which the IMF is not well represented.

\begin{figure}[t]
\includegraphics[width=0.48\textwidth]{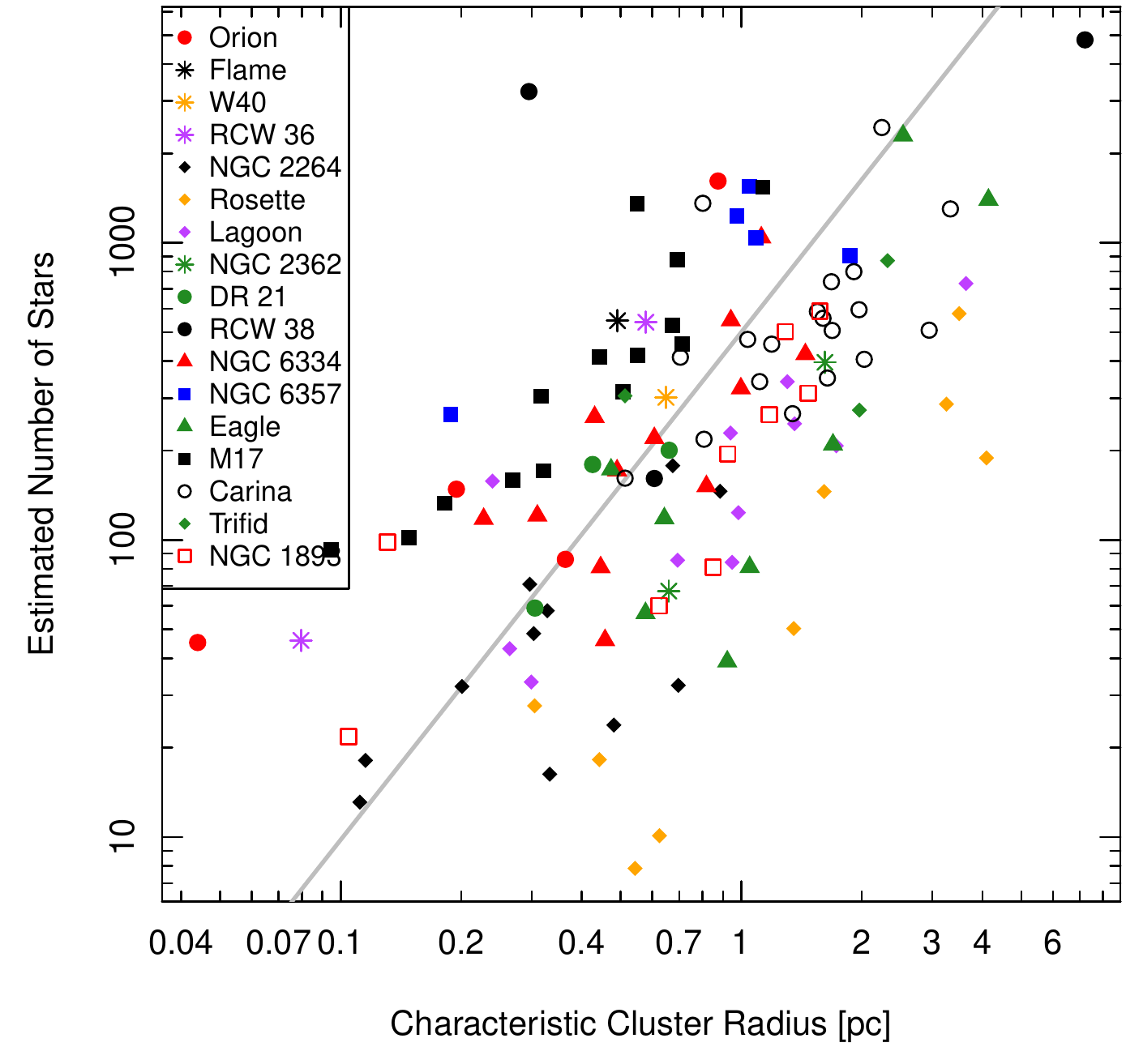}
\caption{Cluster mass as a function of cluster radius for the data by Kuhn et al. (2014) for all the different star forming regions they considered. The lines display the fit to  the Lada \& Lada (2003) data from Fig.\,1. }
\label{fig:cluster_mass_radius_sum}
\end{figure}

More important than the difference in actual fit values, is the difference in the interpretation between Adams et al. (2006) and Pfalzner (2011). Adams et al. (2006) take the relation as an initial condition for their simulations. By contrast, Pfalzner et al. (2011) interpret the slope as evolutionary tracks along which clusters develop. In this interpretation most clusters leave these tracks early on as low-mass clusters, while only a few contain sufficient gas to develop into high-mass clusters.

In both papers the data have a large scatter around the regression line. Here we will test whether the relations continues to hold for larger, more recent samples with samples selected using different wavelengths. In the following we test whether these additional data confirm the existence of a $M_c-R_c$ relation.

We start by looking at the data in Kumar et al. (2006). Their approach to cluster selection is somewhat different to that of Carpenter (2001) and Lada \& Lada (2003), as they do not start out with a list of known clusters. 
Instead they systematically searched for embedded clusters around  the high-mass protostellar candidates listed by Molinari et al. (1996) and Sridharan et al. (2001). They identify the potential clusters as star count density enhancements above the mean background level using the existing $K$-band observations of the 2MASS database. As \mbox{Fig.\,1c)} shows, the data by Kumar et al. (2006) are limited by a similar slope but shifted to lower values and contain additional data points below the line representing it. 
Here the open symbols indicate clusters at large distances (d $>$ 3\,kpc), where mass and radius determination is naturally more difficult. The scatter is significantly reduced by excluding these data points. Only for small cluster masses the wider scatter remains (see Section\,2.3 for a discussion of low-mass clusters).

Two objects diverge significantly from the general $M_c/R_c$ relation, IRAS 06061+2151 and
IRAC 20406+4555. The object IRAC 20406+4555 could have significant uncertainty in size, distance, or number of stars due to its estimated distance of 11.9\,kpc.
The only object that definitely lies above the curve is IRAS 06061+2151, which is nearer, at 2\,kpc, and should be well enough resolved being located in the Galactic anti-centre. This cluster seems to deviate from the ``norma'' star-formation mechanism, which we discuss in Section\,5.

There are several possible explanations for the Kumar et al.  data points lying systematically at lower values: One reason might be that they focus on high-mass star forming regions, where the instantaneous SFE is significantly lower ($\approx$ 10\%) for the high-mass clumps (Johnston et al. 2009). This would mean that they have not formed a large fraction of the stars yet that we would expect for the final cluster. 
In addition, they define the cluster radius as the area enclosed by the contour in stellar density that gives them a cluster detection with 2$\Sigma$-noise level in the stellar density measurement. This is the radius that encloses all mass and is therefore larger than the half-mass radius producing a systematic shift to the right in Fig.\,1c. Any of these reasons, and perhaps a combination of these, may lead to the shift of the curve. Alternatively, it could indicate that the relation actually defines the upper limit for the detection of clusters as surface enhancements (see Section\,2.4). 

\begin{figure*}[t]
\includegraphics[width=0.8\textwidth]{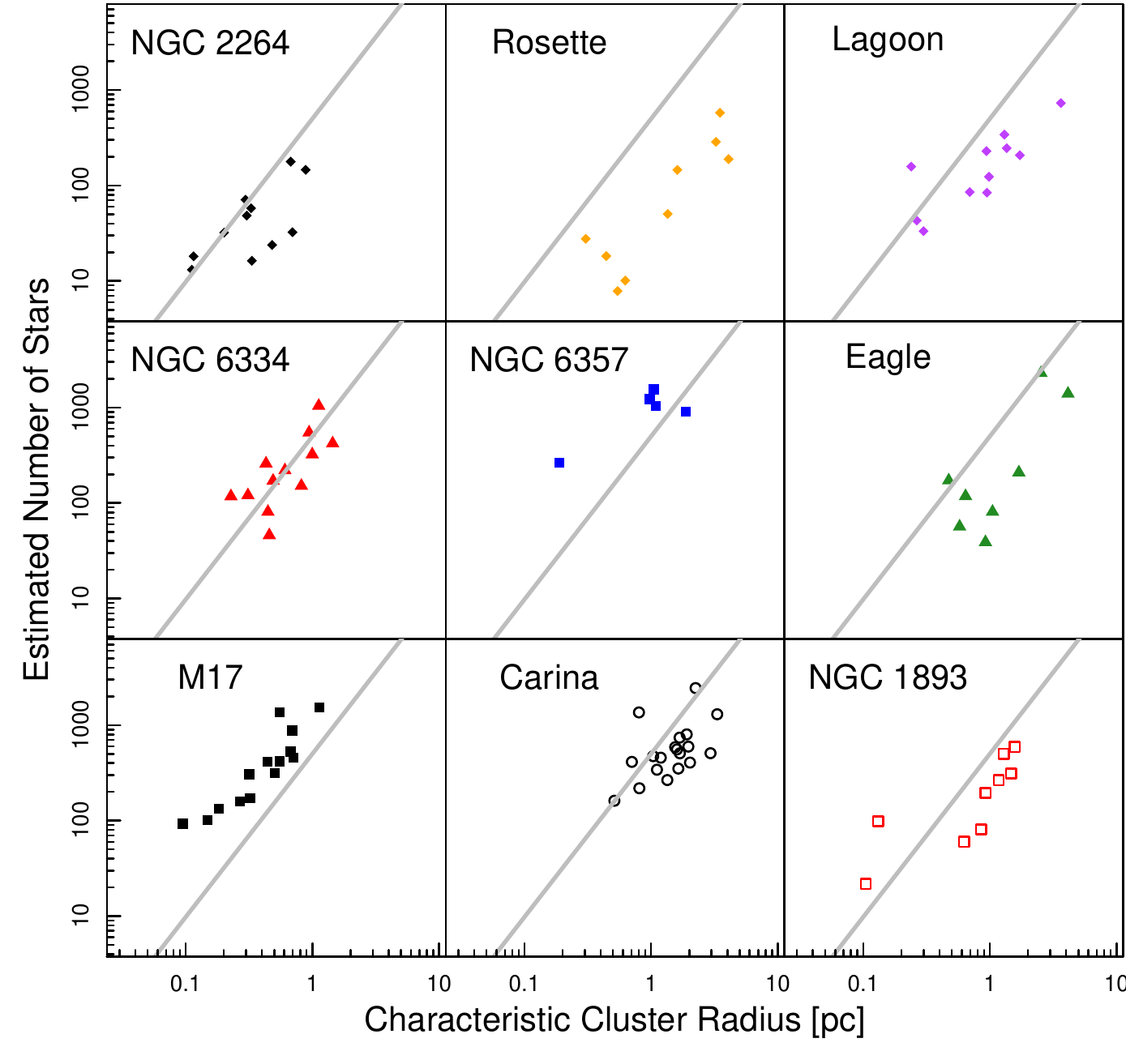}
\caption{Cluster mass as a function of cluster radius for the  data by Kuhn et al. (2014), but separated into the different star forming regions. Here, only star forming regions containing at least 4 subcluster were considered.}
\label{fig:cluster_mas}
\end{figure*}

In summary, it can be said that all data in the $K$-band show a $M_c/R_c$ relation with a similar slope.
The relative shift of the slopes to each other could either be caused by selection effects or different definitions of cluster size.

\subsection{X-ray observations}

A second class of observations are X-ray selected samples. Kuhn et al. (2014) analysed the cluster properties in 17 massive star forming regions (SFRs) using a selection of cluster members detected in the X-ray and infrared in the MYStIX sample (Feigelson et al. 2013; Broos et al. 2013). Compared to IR-only studies, X-ray sources have the advantage of improving the detection of stars at the centres of dense clusters and avoiding contamination from field stars. Nevertheless, these X-ray samples are also not complete, and completeness varies from region to region based on telescope exposure time, distance, and absorption.

Kuhn et al. assume that the star clusters have a isothermal ellipsoid distribution function, and determine the semi-major and semi-minor axis of the core ellipse from their data using a ``Mixture Model." The number of stars is determined within four times the size of the core. They use this cut-off because the projected half-mass radii of the subclusters are difficult to measure. 

The intrinsic numbers of stars in a subcluster may be estimated from the incomplete, observed star counts. In the MYStIX catalogs, typically 16\% of the total population of stars (down to 0.1~$M_\odot$) is observed. However, the the X-ray luminosities of the observed stars can be used to extrapolate the total number of stars based on a universal X-ray luminosity function \citep{FeigelsonGetman05}. The assumption of a universal X-ray luminosity function has been used for this purpose in a number of studies \citep[e.g.,][]{Getman06,Broos07,Wang07,Wang08,Wang09,Wang10,Kuhn10,Wolk11,Wang11,Mucciarelli11,Caramazza12,Getman12}. \citet{2015arXiv150105300K} find agreement within a factor of $\sim$3 with extrapolations based on the IMF, and report the corrected stellar counts in their Table\,2.

Figure\,2 shows the number of stars as a function of the subclusters' reported radius\footnote{They use a radius, four times the size of the subcluster core radius, as a characteristic radius of the subcluster.} 
for the MYStIX sample. Over several orders of magnitudes the distribution of subcluster sizes and radii follow the trend from Lada \& Lada (2003), but with large amounts of scatter. The precise definition of cluster radius is likely to provide an uncertainty in vertical shift for the points, since the four times the core radius, as used by Kuhn et al.\ (2014), is likely to underestimate the half-mass radius (cf. Hillenbrand \& Hartmann 1998). Nevertheless, the slope of the distribution matches the earlier results.

In Fig.\,3 instead of plotting their entire sample in one graph, we show each star forming region (SFR) separately. Here we considered only
star forming regions that contain at least 4 clusters. It can be seen that for each of the different star forming regions a correlation between mass and radius exists and that they are well represented by Eq.\,4. However, some cases have a vertical shift: the subclusters in Rosette lie below the line while in M17 they lie above the line. Overall, the scatter for these individual regions is slightly less than when all regions are combined.  It is worth noticing that the scatter in Fig.\,2 seems to come predominantly from SFRs with few clusters.

Note, \citet{2015arXiv150705653K} find a regression line of the form $\log N = 2.5 + 1.7 \log R$  for the combined data from the the SFRs they investigated. There, they interpret the density vs.\ size relation, which is slightly less steep than would be expected for expanding subclusters with a conserved number of stars, as possible evidence for subcluster mergers. However, we suggest that this relation may be inherited from the mass--radius relation of molecular clouds.

\subsection{Extension to low- and high-mass clusters}

\begin{figure}[t]
\includegraphics[width=0.5\textwidth]{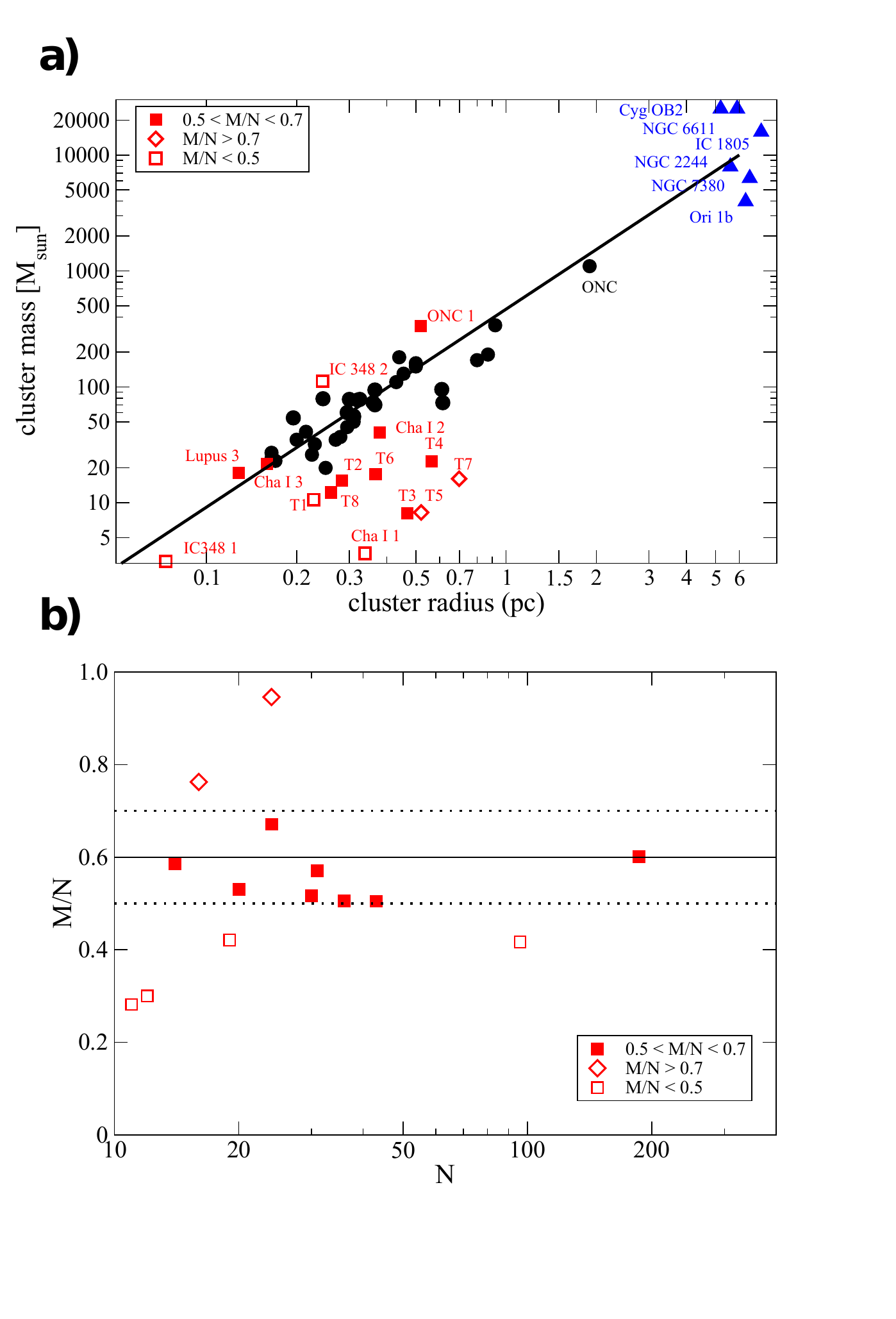}
\caption{a) Same as figure 1b (black circles), but also containing the clusters from Kirk et al. (red squares) and the  OB associations (or extended clusters) from Pfalzner \& Kaczmarek (2013) (blue triangles).  b) Average stellar mass as a function of number of detected stars for the data from Kirk et al. The full line indicates the tyical value of themean cluster mass $\bar{m}$=0.6, the dashed lines deviations of $\pm$0.1 from this value. The open symbols indicate clusters for which \mbox{0.5 $< \bar{m} <$ 0.7} is not fulfilled.}
\end{figure}

Studying clusters of higher mass ($M_c \geq $ 10$^{4}$\,\Msun), we are confronted with the problem that
currently there is no cluster in this mass range known that is still embedded in its natal cloud.  The reason is probably two-fold: first, massive clusters are rarer than low-mass clusters, so that the sample size is smaller; second, massive clusters contain many massive stars that expel the gas much more quickly, so that the embedded phase of massive clusters is relative short ($\approx$ 1\,Myr).
Therefore we consider only massive clusters younger than 4\,Myr that have just lost their gas and might have already started to expand (see Pfalzner \& Kaczmarek 2013, Pfalzner et al. 2014). 

It is interesting to note that if one includes the massive young clusters/associations ($<$ 4\,Myr) from the solar neighbourhood (for a list see for example, Pfalzner 2009) to Fig.\,1b) they are positioned at the extension of Eq.\,4 (see Fig.\,4a). These clusters are about an order of magnitude more massive than the ONC, which is the most massive cluster in the Lada \& Lada (2003) sample. Thus it seems that the same formation mechanism is at work for clusters with a range in total stellar membership from a few dozen stars to tens-of-thousands (an order of magnitude more stars than the ONC).

We can also extend our analysis to lower-mass clusters, although it is not clear when clusters stop being clusters and are just high order multiple systems. Even so, Kirk et al. (2012) show that these small stellar groups share many properties with their higher mass counterparts. In Fig.\,4b the data for the low-mass clusters given by Kirk \& Myers (2011) have been added. These data show a broad scatter from 0.3 -- 1.0 around the anticipated relation. One reason is probably again that with such a low number of stars the IMF is not sampled well.
Figure\,4b shows the cluster mass as a function of the number of stars in this sample. It can be seen that about half of the sample clusters are outside the range 0.5 $< \bar{m} <$ 0.7.
These clusters are indicated as open symbols in Fig.\,4a. Among them are some of the clusters that deviate most from the relation like, T5, T7 and Cha I 1. However, even if one considers only the clusters that represent the IMF reasonably well, there still remains a wide scatter.

Measuring sizes for clusters with a small number of members is much more uncertain than for larger systems.  For consistency with the other datasets plotted, we estimate the size of the Kirk \& Myers (2011; hereafter KM 11) clusters based on the effective radius needed to encompass half of the total cluster mass. The smallest clusters have less than 20 members, suggesting that statistical uncertainties will be much higher for these systems than for much larger clusters, although this would be expected to produce a greater scatter, but not a shift off of the main relationship.    The typically larger sizes for a given mass of the KM11 clusters may, however, be explained by observational sensitivity.  All of the clusters in KM11 except for the ONC, were based on very deep stellar catalogs ($\approx$0.02\,\Msun), and may include more low-mass sources than the other clusters presented.  Since KM11 find a tendency for the most massive cluster members to be centrally located, a sensitivity to lower mass stars could be expected to typically increase the estimated cluster radius, relative to what  a shallower catalog of the same cluster sources would imply.
 
Another reason might be that at the low-mass end of the cluster formation somewhere the transition to isolated star formation occur.  Isolated star formation does not necessarily mean that only one star forms but it could be a binary or even a higher order system. Here the question arises where we speak of the formation of a multiple system and where we think of it as a cluster.  With systems like IC 348 1 and ChaI 1, both having masses $<$5\,\Msun, we might already have reached this transition region. 

\subsection{Systematic Effects}

The different methods used to determine radius and mass in the different studies is a general problem in this context. In some studies the clusters size istaken as the maximum extent of the cluster, in others the half-mass  or half-light radius or as in Kuhn et al. (2014) even four times the core radius.
Similarly, membership and therefore mass definition, varies from study to study. In addition, some studies just give the number of observed stars, while some correct for the IMF.

Given the diversity in radius and mass definitions in this context, the similarity of the obtained mass-radius relations is actually amazing. Why do the differences in radius and mass definition {\em not} influence this relation strongly?  Is the slope of $\gamma \sim$1.7$\pm$ 0.2  just a coincidence or real?

The fact that one mostly obtains such a slope despite these differences hints at a self similar mass-radius development throughout. The likely key property is the stellar number and mass distribution within the clusters themselves. The structure of a spherically symmetric cluster would be dependent on the distance to cluster centre, $r$, as $\propto r^\alpha$ with $\alpha$ between \mbox{-1.5} and \mbox{-2}. The cluster surface-density profile assumed by Kuhn (2014) approaches an $\alpha=-2$ power-law with increasing $r$. If  we assume a unique mass-radius relation exists, then different mass radius definitions basically move the data points, more or less, along the line of the mass radius-relation. 
Thus it is the similarity in slope between the mass radius-relation and the stellar mass distribution within the cluster that makes its detection more or less independent of which measure is actually used.

Nevertheless, future work should aim at a standardized way of determining the radii and masses of clusters to remove this uncertainty from the data samples. Kuhn et al. (2014) try to do this, but their method requires that a model be assumed. Despite these uncertainties the observations all strongly indicate that at least for still embedded clusters in the solar neighbourhood it holds that more massive cluster are more extended. 

There is  an additional problem with determining the cluster size, membership and mass. These properties are determined by comparing the surface density to that of the field. The cluster extent is then defined as the radius where the surface density drops below that of the field population.
The real extent of the cluster might exceed this area as the outer parts of the cluster might have a
lower density than the projected field surface density.  Depending on the distribution of stellar mass in the cluster mass profile this might shift the curve. If all clusters are relatively massive and reside in areas of similar field surface density this effect might just lead to a slight shift. The problem is likely more severe for low mass clusters than for more massive ones. Similarly X-ray studies are contaminated by AGNs. For the Chandra studies the small field of view (17'x17') means that the outskirts of the clusters are simply not mapped.

The fact, that very few of the investigated clusters reside significantly below the mass-radius relation might be simply a selection effect. Clusters that might exist below the mass-radius relation  would have a relatively low mass for their size resulting in a very low surface density. As such they might not be detectable as enhancements above the surface density of the field population (Pfalzner et al. 2015).
GAIA observation will possibly allow to answer the question whether there are no clusters in this part of the parameter space and whether the clusters might extend further than we previously thought. However, GAIA will observe mainly in the Galactic Plane, where clusters at distances greater than a few kpc may be obscured by tens of magnitudes of V-band extinction.

More surprising is that we do not find any clusters,  apart from perhaps the two exceptions in the Kumar et al. data (discussion see below) that are situated significantly above this line, but are below that of the compact clusters of the Galactic Centre and the spiral arms. Clusters with, for example, \mbox{$r_{hm}$ = 0.3\,pc} and $M_c$ = 1000\,\Msun\ seem to not exist or are at least rare. Clusters above the $M-R$ relation should actually be more easily detectable than the other clusters in the sample due to their higher surface densities. It seems that clusters with such high volume/surface densities either pass this parameter space very quickly, are very rare, or simply do not exist, at least not in the solar neighbourhood.

\section{Gas Clumps}

\begin{figure}[t]
\includegraphics[width=0.43\textwidth]{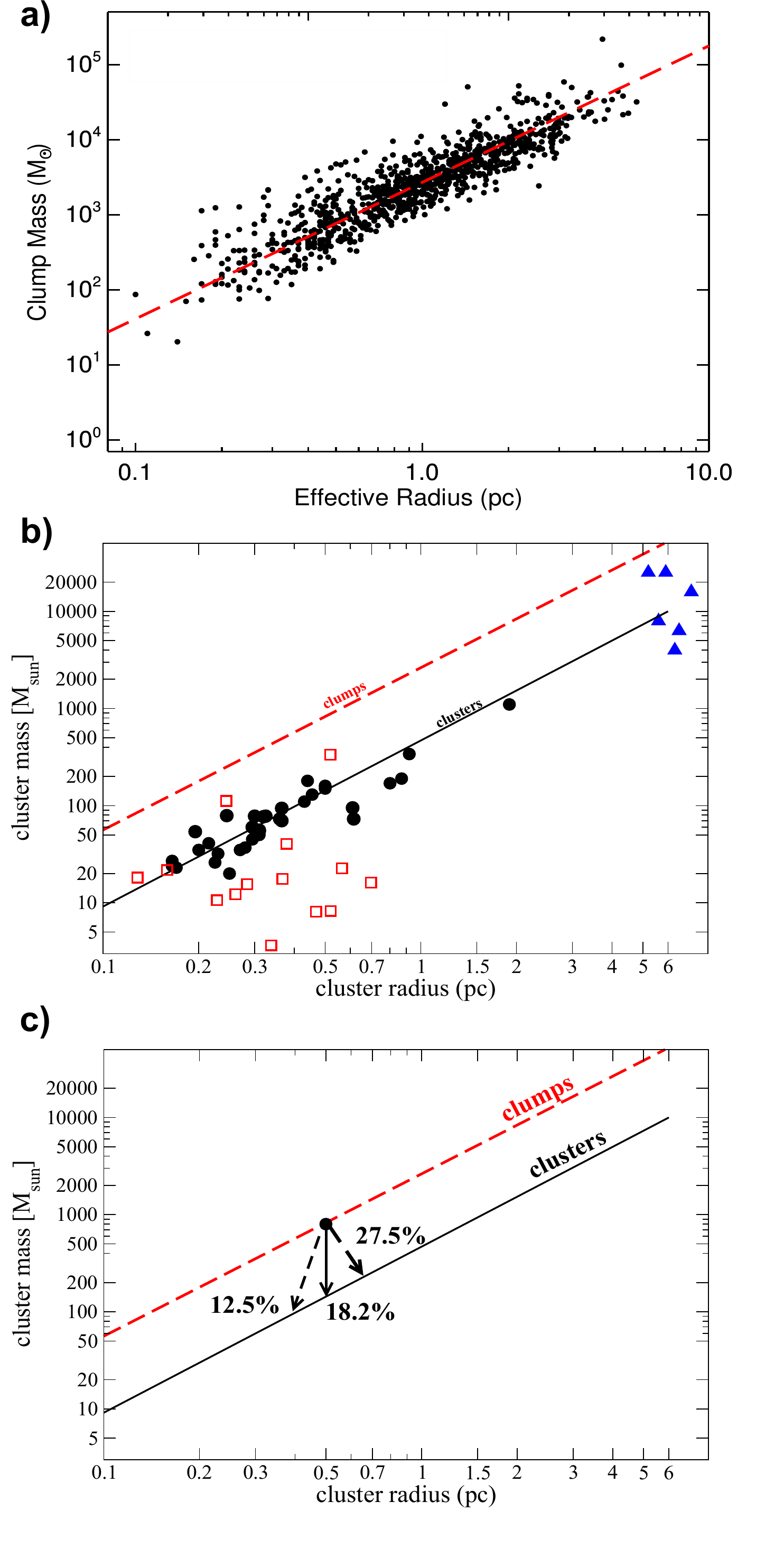}
\caption{Clump masses as a function of clump radius. a) The upper panel shows the distribution of massive star forming clumps identified by cross-matching dense clumps with methanol masers, massive young stellar objects and compact HII regions, all of which are excellent tracers of embedded massive stars and their associated proto-clusters (Urquhart et al. 2014). The dashed red line shows the results of a power-law fit to the clump size-mass distribution. b) Comparison
between the clump data by Urquhart et al.  and the cluster data of Fig. 4a, c)  Possible range of star formation efficiencies derived from the mass-radius relations for gas clumps and embedded star clusters in the solar neighbourhood. The red-dashed line indicates in all three plots the fit of Urquhart et al. 2014 and the black full line the fit to the Lada \& Lada 2003 data. The symbols are the same as in Fig. 1 and 3.}
\label{fig:cluster_mass_radius_sum}
\end{figure}

As a next step we investigate  compact regions of dense material within giant molecular clouds usually referred to as clumps, and  exclusively investigate the high-mass end of clumps with $M_{clump} >$ 20\,\Msun.  These clumps are sites of star formation,
and many will collapse to form star clusters. In this study, we will concentrate on those clumps that have the potential to form clusters of at least a dozen members. Gas clumps that form only one or a few stars are usually referred to as cores. Although the core-star connection is a very relevant subject, as many recent studies have shown (for example, Alves et al. 2007, Lada et al. 2008), we will exclude it from the current investigation.

Here we take the data from the recent survey by
Urquhart et al. (2014), who presented a study of every ATLASGAL clump associated with massive star formation identified across the inner Galaxy. The clump masses were estimated assuming an isothermal temperature of 20\,K using the method of Hildebrand (1983) and the radius is determined from the standard deviation of the pixel weighted dust emission distribution multiplied by a factor of 2.4, which relates the standard deviation to the effective radius of the clump (see Rosolowsky et al. 2010 and Contreras et al. 2013 for more details). They find a strong correlation between these two parameters,
which they fit as $\log (M_{clump}) $ = 3.42 $\pm$ 0.01 + (1.67 $\pm$ 0.025 $\times \log (R_{clump})$) (see Fig.\,5a for an adapted version of their Fig.\,24). In Fig.\,5b we compare the slopes of clusters to those of clumps in the mass-radius plane. Furthermore,  the slopes for the mass relation for star clusters (1.7$\pm$ 0.2) is, within the errors, the same as that of the gas clumps  (1.67 $\pm$ 0.025). This means there is likely to be a correlation between the cloud mass and the cluster mass functions, perhaps analogues. 

When comparing the slopes of the mass--size relation for clusters and clumps one must keep in mind, that
recent studies (Smith et al. 2008, Pineda et al. 2009, Reid et al. 2010, Curtis \& Richer 2010, Shetty et al. 2010,  Ward et al. 2012) have pointed out the limitations of current observing techniques. They show that it is difficulty to identify and quantify properties of these clumps that reflect the true nature of these star forming regions. One of the problems is that one has to deal with projection effects.

An additional problem is that clump masses and radii (like the star cluster masses and radii) are determined in different studies in different ways. One obvious component is the sensitivity of the instrument used. The resolution of the instrument also has an important effect, because for more distant regions (e.g., for MYStIX regions at 2--3\,kpc) a clump that has a radius $<$0.1 pc may be only barely resolved, and any substructure may be lost. However, as long as the masses and radii for a given sample are measured in a consistent way the trend (i.e., slope) is invariant to the method used as it affects all sources systematically --- they simply shift a little but the overall distribution remains the same.
Using the Urquhart et al. (2014) sample has the advantage of such a homogeneity in size and mass determination. We also find that the slope is not biased by distances as the same result is obtained using a distance-limited sample. So at least as far as the clump distribution is concerned the distribution and slope is robust. As discussed in Section\,2, this homogeneity of the sample is not to the same degree given for the cluster sample.

Relating clump to cluster properties is the next difficulty. Here one has to be sure to detect the entire projected area of the region in which the gas density is high enough that star formation takes place, but exclude gas areas that do not contribute to the star formation process. One constraint is that the clouds have to be in molecular form to be able to form stars, therefore one needs $A_V > 1$\,mag, as otherwise H$_2$ is dissociated. This means that, in projection, this outer edge has to have at least  $A_V > 2$mag. If we take this as the outer edge of the star forming region of clouds, the average column density in clouds  in the solar neighbourhood is \mbox{$\sim$ 4\,mag} which is equivalent to 80\,\Msun\ pc$^{-2}$ (Lombardi et al. 2010) and Andre et al. (2010) find that star formation starts at a threshold of A$_V\sim$8\,mag.  The ATLASGAL 5$\sigma$ sensitivity is $\sim$0.25\,Jy\,beam$^{-1}$, which corresponds to column densities of $3.6\times10^{21}$\,cm$^{-2}$ for dust temperatures of 30\,K and $2\times10^{22}$\,cm$^{-2}$ for dust temperatures of 10\,K, equivalent to 4 to 21\,mag of visual extinction (Schuller et al. 2009).
If the clusters are measured in a consistent way, and not necessarily the same way as the clumps, then both slopes will be robust but the offset between them is less reliable.

Keeping all these issues in mind, we assume that the two nearly identical slopes in the mass radius plane of the gas dumps and the clusters are real. An obvious idea is to use these two lines to determine the SFE (see Fig.\,5c). First we assume stars form within the entire clump radius, meaning that the clump and the cluster radii are the same (this corresponds to a vertical line in Fig.\,5c). In this case, we obtain an average SFE for the embedded clusters in the solar neighbourhood of 18\%. This is in agreement with measurements of SFEs of single embedded clusters, which lie roughly in the range 10\%-30\% (for example, Lada \& Lada 2003).  

As pointed out above one must be careful with this approach; as discussed in Section\,2.4 the real extent of the cluster and therefore the original extent of the gas clump might exceed this
observed area. Depending on the distribution of stellar mass in the cluster mass profile this might shift the curves relative to each other. If we take these potential errors into account, the average SFE could be in the
range 12\% $<$ SFE $<$ 28\%. This is again in the range of SFEs one obtains when comparing the stellar mass to the gas mass in induvidual clusters. However, one has to keep in mind that incomplete sampling of the IMF is a big source of uncertainty in these studies. 

For the star clusters considered here SFE $\le$ 30\% and expel $>$ 70\% of the initial clump mass at the end of the star formation process. As a consequence the majority of the cluster members (80\%-90\%) become unbound and only a small
remnant cluster containing 10\%-20\% of the original stellar members remains a bound entity (Hills 1980, Lada et al. 1984, Geyer \& Burkert 2001, Kroupa et al. 2001,  Bastian \& Goodwin 2006 Baumgardt \& Kroupa 2007, Pfalzner \& Kaczmarek 2013).
The 80\%-90\% of stars that become unbound join the field star population. This means all of the clusters in the  various samples investigated here will not become long-lived clusters but most of their stars will join the field population. Only a fraction of the stars will remain bound in a remnant cluster for some tens of Myr.  Even relatively massive clusters like the ONC will face this fate
(Pfalzner et al. 2015) and presumably the smallest cluster systems will in fact be entirely dissolved on a much shorter timescale.

Is the mass-radius relation in any way connected to a threshold gas
column density above which star formation takes place?  Lombardi et
al. (2010) give $A_V$ = 10.5 mag or \mbox{220\,\Msun\,pc$^{-2}$}
threshold.  Interestingly many of the clusters lie well above this
limit as do most of the ATLASGAL clumps (typical clump masses are
  2-$3\times 10^3$\,\Msun; Urquhart et al. 2014). This means that the
star clusters form only in those areas of the clouds where the column
density is much higher than this limiting value.  Interestingly,
Parmetier et al. (2010) found that a volume density thresholds overall
for star formation implies a mass-size thresholds for massive star
formation.

Kauffmann et al. (2013) find a threshold for clouds to become
  self-gravitating at $M\approx{}770\,\Msun\,(R_{clump}[pc])^{1.63}$
  (their Eqs.~1 and 12, combined with a virial parameter $\alpha=1$).
This limit corresponds much better than the Lombardi et al.\ star
formation limit to what we see in the cluster mass-radius
relation. Kauffmann et al. (2013) interpret it as the limit where
competitive accretion becomes important. This is a consequence of
  the linewidth--size relation of ``turbulent'' random gas motions
  within molecular clouds, as for example described by the Larson
  (1981) relations. See Kauffmann et al. (2013) for a recent
  compilation of linewidth--size data in dense gas. However, at least
the small clusters, like the Taurus clusters in Kirk \& Myers (2011),
have too few members and are too sparse to really fit the competitive
accretion picture.

Recently a number of efforts advocate Kennicutt-Schmidt like laws within individual Galactic star-forming regions, e.g. Gutermuth et al. 2010. These efforts have been criticized (e.g. Lada et al. 2013) as an artifact of statistical effects rather than being a real astrophysical phenomena. 
Possibly our method of looking at entire clumps and entire clusters is likely to be more robust, than looking at smoothed surface density maps of stars and clouds. 

\section{Comparison with theoretical models 
and simulations}

\begin{figure}[t]
\includegraphics[width=0.5\textwidth]{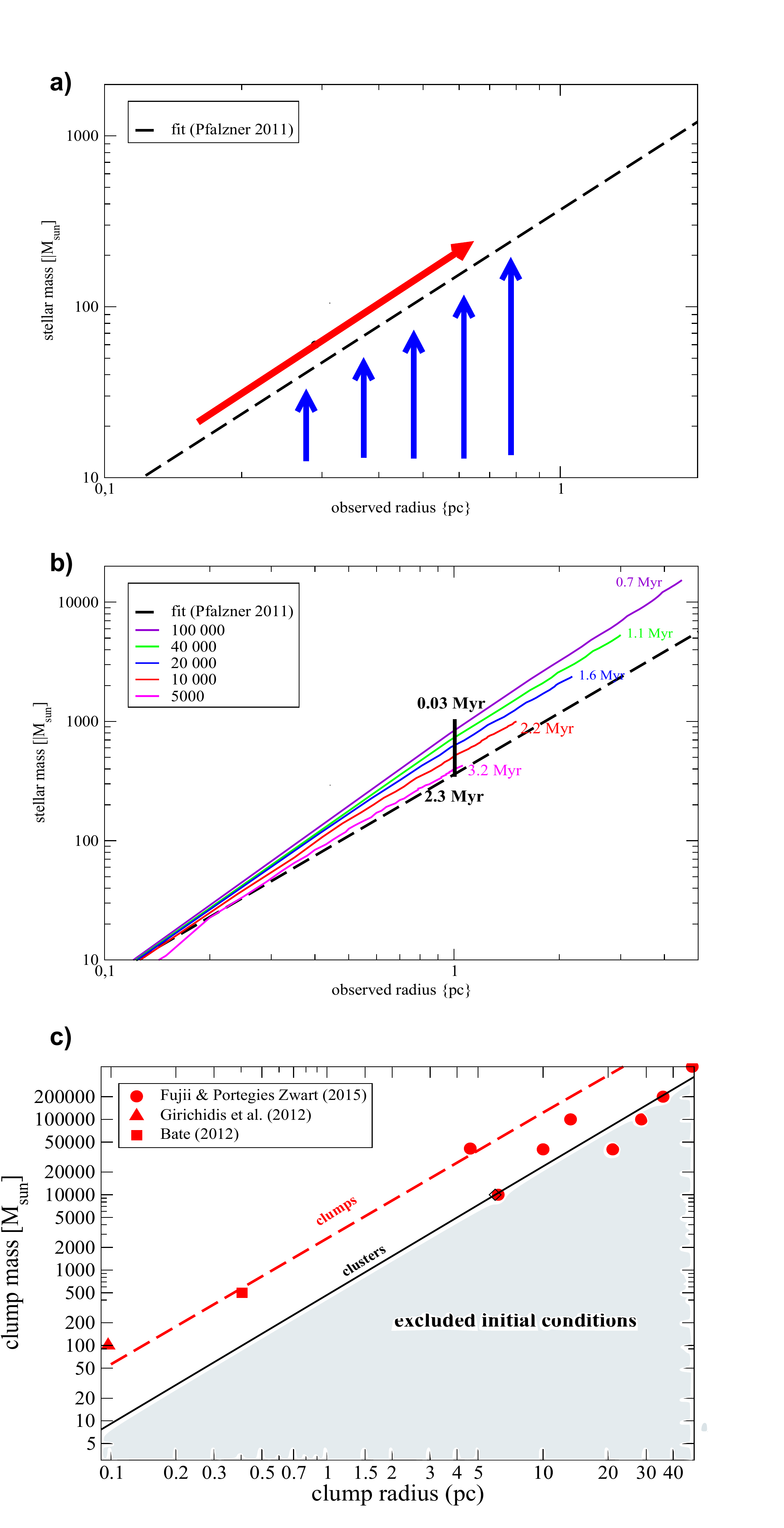}
\caption{a) Schematic picture of cluster formation driven  by hierarchical clump structure 
b) Mass size development for different clump masses in the free-fall regulated scenario (adapted from Pfalzner et al. 2014). c) initial conditions used in various numerical studies of clustered star formation. }
\label{fig:clump_cluster_schematics}
\end{figure}

\subsection{Interpretation of the observational data}

Before we test to what extent the existing theoretical models of star cluster formation reproduce the mass-radius relation, we want to first have a look at how the mass-radius relation can be interpreted as a formation history. There are several ways one can interpret this mass-size relation:
\begin{itemize}
\item The easiest assumption (illustrated by the blue arrows in  Fig.\,6a) is that clumps of different radius form clusters of more or less the same radius. If they do that simultaneously over the entire area, clusters would ``grow'' from low-mass clusters to their final mass without changing their size, and their size would be determined by the extent of the clump. 
\item
This picture does not take into account that the lower gas density in the outer regions of the gas clump
leads to longer free-fall times in these areas. Longer free-fall times could result in slower star formation in the cluster outskirts than the centre. This slower star formation in the outskirts would appear as an increase of the cluster size and mass with age (illustrated by the red arrow in Fig.\,6a). The observation Kuhn et al. (2015a) support this scenario in that they find that the low-mass clusters are on average younger than the high-mass clusters and Kuhn et al. (2015b) provide free-fall times for the MYStIX subclusters. These are generally less than the cluster ages.
\item
As mentioned in Section\,2.4, the cluster size and membership is currently largely determined by
comparing the cluster surface density to the background density. This always leads to underestimating the cluster extent and this effect is stronger in areas of higher background density. Due to their lower
surface density, low-mass clusters are generally more affected by this situation than more massive clusters.
\end{itemize}

All these effects and different interpretations have to be considered when comparing the theoreitcal models with the observations.

\subsection{Theoretical models}

There exist many different analytical and numerical models that describe the formation of star clusters.
Here we will restrict ourselves to several examples that we discuss in more detail. However, every model of clustered star formation that is representative of the solar neighbourhood needs to provide an explanation for the observed mass-radius relation. We will see below that some of the standard models of clustered star formation are not representative of the clusters one finds in the solar neighbourhood
but rather of  the type of clusters one finds in the higher density regions of the spiral arms or the Galactic centre (see Section\,5)  or is suitable to point out the differences between both types of clustered star formation.
 
One analytical model for the formation of clusters is that of scale-free hierarchical star formation. In this model high-velocity turbulent gas form large scale structures, whereas low-velocity compression forms small clumps (Elmegreen 2008, Kruijssen et al. 2012). A direct consequence of this model would be that young stellar groups are hierarchically clustered, with the large stellar complexes at the largest scale and the OB associations and subgroups at the smallest scales (e.g. Efremov 1995). At face value this model contradicts the observed mass-radius relation, because the OB associations actually have relatively large sizes. However, the observed OB associtions are usually older than the clusters treated here. In addition  this model is probably not applicable to the solar neighbourhood itself, but is more suitable to explain the difference between the compact clusters found in the high-density regions and the extended clusters one finds in the solar neighbourhood.

A general problem in this context is that very few of the analytical models explicitely treat these
embedded clusters that are typical for the solar neighbourhood, which probably largely dissolve after the gas expulsion process.  Much more attention is given to clusters that survive this developmental stage more or less intact and have the potential to develop in long-lived open clusters or even globular clusters. This is despite the fact that most star formation (80\%-90\%) in the Milky Way happens in clusters like those typical for the solar neighbourhood (Bastian 2013). 

Even fewer studies take some mass-radius relation into account. The first one that tried to find a description of this process was Pfalzner (2011). There it is assumed that star formation starts in the central area of the clump, continues in the central area while the star formation front moves outwards in a medium
with an $r^{−2}$-gas density dependence. Although the mass-radius relation is reproduced in this model, this is done in a more or less descriptive manner and the underlying physical process is not quite clear. More recently a model has been proposed that is based on the dependence of the free-fall time on the gas density (Parmentier \& Pfalzner 2013, Pfalzner et al. 2014). This model reproduces the
mass-radius relation and takes into account the differences between observed and real cluster properties (Fig.6b). It is the first model that reproduces the mass-radius relation from first principles. However,
so far it is mainly aimed at explaining the formation of high-mass clusters and therefore assumes a relatively large clump size. The effect of an entire spectrum of clump sizes, as indicated by the observation, has to be the next step in this context. 

\subsection{Simulations}

The major problem when simulating clustered star formation from an ab initio scenario is that the initial conditions are not well
constrained by observations. As such a number of assumptions concerning the initial density distribution, the corresponding temperature, the chemical composition etc. have to be made. Similarly there is limited information about specific parameters in feedback calculations etc. In addition, any numerical model has to include further approximations due to computational limitations, for example concerning the spatial or temporal resolution, like for example, sink particles sizes etc. Girichidis et al. (2011) demonstrated that the outcome of such simulations is very sensitive to the chosen simulation parameters.

Therefore there always remains some uncertainty to which extent the numerical models really represent
the actual physical situations. Above described mass-radius criteria could provide a test in how far
the models are representative for the real physical situation. It would therefore be interesting to see to what extent current simulations can reproduce this relation. 

The results of simulations of clustered star formation  usually compare properties like the initial mass function and multiplicity with observed properties (e.g., Bate 2012). It is not clear to what degree these properties are influenced by the cluster environment or whether they are the same in isolated formation.
So far, properties specific to clustered star formation have been less investigated in these simulations.
It would be relatively straightforward to check whether these simulations reproduce the observed mass-radius relation described here.  

The first step is to look at the initial conditions in these simulations --- meaning the initial clump mass and radius. We have seen that the mass-size relation, not only for the clusters but also for the clumps, is  very strigent. If we want to simulate the formation of clusters that are representative for the observed ones, the initial conditions for the simulations have to be chosen in such a way that they correspond to the mass-radius relation of the clumps. In Fig. 6c the initial conditions for some star formation simulations are shown. It can be seen that the initial conditions in the simulations are mostly representative for the observed clumps. The exception are some of the high-mass clusters where the sizes are too large for the choosen masses. The simulations that start with realistic initial conditions could now be tested whether they lead to the observed mass-size relation.

Unfortunately, these simulations are still restricted to relatively small cluster masses due to computational limitations. However, the parameter space of clusters up to 100\,\Msun\ could already be tested.  If we take the most recent cluster-forming simulations by Bate (2012) as an example, the final stellar mass in their clusters lies between 28.4\,\Msun\ and 92.1\,\Msun. We would expect a cluster size of approximately 0.12\,pc and 0.35\,pc, respectively. However, the advantage would be that one can not only compare the sizes of the end product, but the entire developmental process.
Reproducing the observed mass-size relationship in simulations would provide strong constraints on the initial conditions and should be a fundamental requirement.

Girichidis et al. (2012) simulate  additional  parameters that are specific for clusters like the virial state of the cluster and degree of mass segregation for a range of simulations. However, they suffer again from the numerical limitations on cluster mass.
All their simulations assumed a total cloud mass of \mbox{100\,\Msun}, thus even for a high  SFE of 30\% the
the resulting clusters would not exceed 30\,\Msun. As we showed above, it is only for clusters with $M_c>$ 30\,\Msun\ that we can varify the existance of the mass-radius relation. 

Another problem in these models is that the simulation volume is often relatively small (for the example of Girichidis et al. it is 0.26\,pc). Thus it cannot be excluded that the size of the simulation area influences the size of the obtained cluster. 

Recently Fujii \& Portegeis Zwart (2015) modelled the formation of massive clusters ($M_c >$ 10$^3$\,\Msun); this is only possible by choosing a much lower resolution. It is currently unclear how senstive clustered star formation is to the processes that happen on small scales. Again, they give masses of the obtained clusters but no size estimates. So it is unclear whether these simulations lead to the required mass-radius relation for clusters or not.

%\begin{figure}[t]
%\includegraphics[width=0.5\textwidth]{higuchi_cloud.pdf}
%\caption{Clump (black squares and cluster masses (triangles) as fiunction of clump radius for the data of Higuchi et al. (2009).  }
%\label{fig:cluster_mass_radius_sum}
%\end{figure}

\begin{figure}[t]
\includegraphics[width=0.5\textwidth]{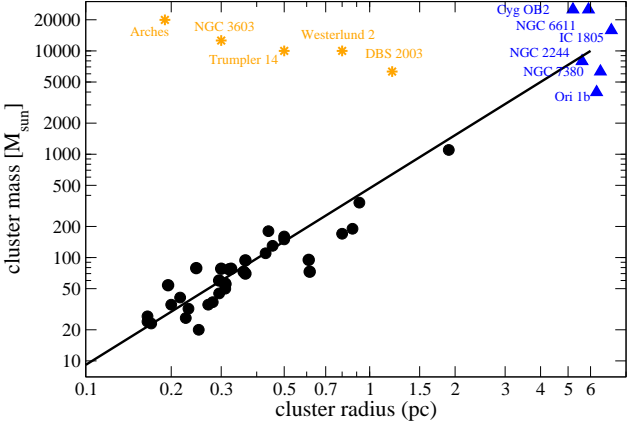}
\caption{Same as Fig.\,5b) but in addition the properties of massive clusters in the spiral arms and the Galactic centre are shown.  }
\label{fig:cluster_mass_radius_sum}

\end{figure}

\section{Discussion}

In the previous sections, we showed that the observed clusters do not fill the entire mass-radius plane but only a very specific parameter space. We have detailed in Section\,2.4 that the absence of  low-mass extended clusters could be caused by detection problems. However, the question arises, why, for example, do no clusters containing a thousand stars and having a half mass radius $<$ 0.4\,pc exist in our samples?  These would be in principle easy to detect due to their high surface density. 

For the data investigated here we can say that clusters above the mass limit $M_{limit} $ 
\be M_{limit} < C_M R_c^{1.7},\ee
do not form in the solar neighbourhood. Here $C$ is of the order of 200 (see Fig.\,6). The possible underlying reason could be that in order to built a cluster that is more compact than those given by the mass-radius relation one would require
considerable energy to compress the gas to the required density.  The energy necessary to compress the gas mass to such high densities could be just too large. Possibly there exists something like a Jeans criterion not only for the formation of single stars, but also for cluster formation.

%Another question is whether the values we obtained so far for cluster mass and radius include actually the entire extend of clusters. The results of recent Gaia-ESO surveys () show  that there is a young, low-mass stellar population spread out over at least several square degrees in the Vela OB2 association. This population of low-mass stars could have originally been part of a cluster around $\gamma$ Velorum that expanded after gas expulsion or formed in a less dense environment that is spread over the whole Vela OB2 region.

The discussion given above has been restricted to the solar neighbourhood, however, the question is whether the
mass-radius relation is of a more universal nature, or not. In Fig.\,7 the massive compact clusters from the Galactic Centre and the spiral arms are added in the diagram of Fig. 5b. These clusters have about the same mass as the most massive clusters in the solar neighbourhood, but are approximately a factor of at least 10 more compact. In addition, several studies have recently suggested that the SFE in these clusters must have been much higher (60\%-70\%) than for the clusters in the solar neighbourhood. These massive compact clusters occupy a distinctively different area in the mass-radius plane than all the embedded clusters in the solar neighbourhood. This seems to indicate that cluster formation in extreme environments like the Galactic centre and the spiral arms proceeds differently to what we know from the solar neighbourhood (Ratheborne et al. 2014).

All the massive clusters used in Fig.\,7 -
compact and extended - have largely lost their natal gas and therefore star formation has more or less ceased. By contrast, the low-mass clusters in the sample illustrated in Fig.\,1 are still forming stars. The question arises whether some of them might develop into such compact clusters? Actually none of them have sufficient gas to do so. If we look again at Fig.\,1c, it might be that IRAS 06061+2151 is actually a candidate for 
a compact cluster precursor, but then it would have to be of a lower final mass. Interestingly IRAS 06061+2151 shows signatures of cloud-cloud collisions as do most most of the compact young clusters in the Galactic centre and the spiral arms (Pfalzner \& Huxor, subm.).

The formation of these compact clusters is much less understood than those of the extended associations. Some star forming regions have been identified as containing possible precursor candidates, but  as yet there are no definite precursors in the form of compact massive embedded clusters. From Fig.\,7 it seems that the formation of massive {\it compact} clusters, like the Arches and NGC 3603, close to the Galactic Centre and in the spiral arms (see section\,4) might be quite different from what happens in the solar neighbourhood (Pfalzner \& Huxor, subm.).

\section{Summary and Conclusion}

In this paper we investigated whether there is a correlation between the masses and sizes of clusters in the solar neighbourhood. The study included also the progenitors of the star forming clusters - massive gas clumps in molecular clouds. Comparing different data compilations we find the following:

\begin{itemize}
\item All investiagted $K$-band and X-ray cluster compilations show a similar dependency of the form
\be M_{c} \propto  R_c^{1.7\pm 0.2}. \nonumber \ee
This is the case despite the different compilations using various methods to determining the relevant values. It is not quite clear whether this relation holds generally, or has to be interpreted as an upper limit,
because lower mass clusters might fall below the detection limit.
\item This relation applies all the way to the massive \mbox{($M_ c >$10$^4$\,\Msun)} young ($<$ 3\,Myr) clusters in the solar neighbourhood that have just finished star formation.
\item For very low mass clusters \mbox{($M_ c <$30\,\Msun)} it is less clear whether this relation still holds. This could be either due to difficulties in determining the cluster masses and radii for such small systems or be caused by a transition from clustered star formation to the isolated formation of higher order systems.
\item The progenitors of star clusters - massive gas clumps - show a very similar mass-radius relation of the form \mbox{$M_{clump}  \propto R_{clump})^{1.67 \pm 0.025}$ (Urquahrt et al. 2014).} Thus the mass-radius relation for embedded clusters and clumps is very similar. 
\item We point out that there are several uncertainties in relating clump to clusters properties. However, at face value one obtains an average SFE of 18\%$^{+9.3}_{-5.7}$ for the embedded clusters in the solar neighbourhood.
\end{itemize}

We conclude that theoretical models --- whether analytical or numerical in nature --- have to reproduce this mass-radius relation in order to be valid descriptions of the processes going on in the solar neighbourhood. We tested some of the existing models of clustered star formation and find that
some of the analytical models actual reproduce the observed mass-radius relation. For the numerical models often only the cluster mass, but not the cluster radius is given, so that it remains for the moment unclear whether they described the observed relation. However, this could be easily tested by the groups that perform these type of simulations.

Using mass-radius relation for testing the validity of clustered star formation models can be only a first step.
Recent observations point at a number of other relations that could be used as tests. We will describe these in a follow-up paper.

\end{document}